\documentstyle[11pt,aaspp4,psfig]{article}
\begin{document}

\def\mathnew{\mathsurround=0pt}
\def\simov#1#2{\lower .5pt\vbox{\baselineskip0pt \lineskip-.5pt
    \ialign{$\mathnew#1\hfil##\hfil$\crcr#2\crcr\sim\crcr}}}
\def\simg{\mathrel{\mathpalette\simov >}}
\def\siml{\mathrel{\mathpalette\simov <}}
\def\Mesz{M\'esz\'aros~}
\def\los{l$.$o$.$s$.$~}

\received{9/28/97}
\accepted{ }
\slugcomment{Ap.J.(Lett.) subm 9/28/97}

\title{Rings in Fireball Afterglows}

\author{A. Panaitescu \& P. M\'esz\'aros }
\affil{Pennsylvania State University, 525 Davey Lab., University Park, PA 16802}

\begin{abstract}

We derive the equation for the surface of equal arrival time of radiation from
a thin relativistic shell interacting with an external medium, representing
the afterglow of a gamma-ray burst produced by a fireball.
Due to the deceleration, these surfaces become distorted 
ellipsoids and, at sufficiently late times, most of the light (either 
bolometric or in a given band) comes from a ring-like region whose width 
depends only on age. We analyze the shape of these surfaces and the radiation 
received from different angles for different dynamic and radiative regimes
and homogeneous or power-law external densities. We calculate angle-integrated
bolometric and fixed frequency fluxes, and we tabulate the most relevant 
parameters that describe the equal arrival time surfaces and the source 
brightness distribution, quantities that are useful for more accurate analytic 
estimates of the afterglow evolution.

\end{abstract}


\section{Introduction}
\label{sec:intro}

The afterglows of gamma-ray bursts (GRB) appear to be well-fitted by
decelerating relativistic fireball models (Tavani, 1997; Vietri, 1997;
Waxman, 1997a; Wijers, Rees \& \Mesz, 1997). This picture (\Mesz \& Rees,
1997), in its simplest form, assumes that the bulk of the radiation comes
from the external blast wave pushed ahead of the fireball with a diminishing
bulk Lorentz factor, which is predicted to produce radiation at wavelengths 
longer than $\gamma$-rays decaying as a power-law in time, in good agreement 
with observations. Two interesting consequences of the deceleration dynamics 
are that most of the late radiation comes from a narrow ring, rather than the
entire visible surface (Waxman, 1997b), and that the usual estimate for
the transverse size of a relativistically expanding cloud under-estimates 
the real one (as we show in \S \ref{sec:rad}).
This has consequences for the apparent expansion rate of the fireball, the 
evolution of scintillation properties of the radio-emitting remnant (Goodman, 
1997; Frail et al, 1997), and the probability of microlensing of GRB afterglows 
(Loeb \& Perna, 1997). The exact shape of the surface, the dimensions 
and the expansion rate depend on the 
dynamic and radiative regimes, as well as on assumed properties of the external 
medium. We present detailed analytical and numerical calculations of these 
properties taking into account the effects of the dynamic and radiative 
efficiency regimes, for both homogeneous and power-law density external media. 
We investigate the equal arrival time surfaces and source apparent width 
evolution, provide estimates of the width of the rings, and present simple 
analytic expressions for both the ``average" longitudinal and transverse sizes, 
in the cases of either bolometric or fixed frequency band observations.
 
\section{Equal Arrival Time Surfaces}
\label{sec:arrival}

Here we derive the equation of the surface that the observer sees at given time 
$T$. For simplicity, we assume that the source of radiation can be approximated 
as a surface (see \S \ref{sec:disc} for a discussion on this approximation). We 
also assume the external medium to be isotropic, but not necessarily 
homogeneous; 
therefore at any lab-frame time $t$ (measured in the center of explosion frame),
the ejecta is spherical. The observer equal-$T$ surface is symmetric with respect
to the line of sight (\los) toward the center of explosion which released the 
relativistic ejecta, therefore its equation is given by two coordinates: a polar
angle $\theta$ measured from this central \los and a radial coordinate $r$. 
In the absence of deceleration, the equal-$T$ surface 
is an ellipsoid (Rees 1966) with semi-major axis $\Gamma^2 \beta cT$ and 
semi-minor axis $\Gamma \beta cT$, where $\Gamma=(1-\beta^2)^{-1/2}$ is the 
constant Lorentz factor of the freely expanding ejecta and $c$ is the speed of 
light. For large $\Gamma$, the ellipsoid is very elongated (high eccentricity). 
When deceleration is present, the shape of the equal-$T$ surface departs from
that of an ellipsoid.

The equation of the equal detector time $T$ surface is 
\begin{equation}
ct - r(t) \cos\theta = cT \;,
\label{surface}
\end{equation}
where $r$ and $t$ are related by $dr=\beta_{sh} c dt$, with $\beta_{sh} c$ the 
speed of the shock that sweeps up the external medium. The Lorentz factor 
$\gamma_{sh}$ of this shock can be approximated (e.g. \Mesz, Rees \& Wijers,
1997) as a power-law in $r$: 
\begin{equation}
\gamma_{sh}=\Gamma_{sh}(r/r_{dec})^{-n}~~~~~;~~~~~ 
                                 n=\frac{3-\alpha}{1+\delta} > 0 \; ,
\label{gamma}
\end{equation}
where $\alpha$ ($<3$) is the index of external gas density power law dependence
($\rho \propto r^{-\alpha}$), $\delta$ describes the dynamics ($\delta=0$ for 
momentum-conserving and $\delta=1$ for energy-conserving evolution), and 
$r_{dec}$ is the deceleration radius, defined as the radius at which
the ejecta has swept up a mass equal to a fraction $\Gamma_0^{-1}$ of its own
mass, $\Gamma_0$ being the initial Lorentz factor of the ejecta. 
Numerical simulations of 
the hydrodynamic interaction between the ejecta and the external medium 
(Panaitescu \& \Mesz, 1997) show that $\gamma_{sh}$ slowly decreases
below the deceleration radius $r_{dec}$ and that 
$\Gamma_{sh} \simeq (2/3) \Gamma_0$. Thus equation (\ref{gamma}) is correct
only for $r > r_{dec}$.

For relativistic shocks, the relationship between $t$ and $r$ has 
a simple form:
\begin{equation}
ct = r + \frac{r_{dec}}{2(2n+1)\,\Gamma_{sh}^2} \left(\frac{r}{r_{dec}} \right)^{2n+1} \;.
\label{time}
\end{equation}
Substituting $t$ from equation (\ref{time}) in equation (\ref{surface}),
the equal-$T$ surface is 
\begin{equation}
\theta = 2 \sin^{-1} \left( \frac{1}{2 \Gamma_{sh}} \sqrt{\frac{\tau}{a} - \frac{a^{2n}}{2n+1}} \right) \; ,
\label{equalT}
\end{equation}
where the reduced variables $a=r/r_{dec}$ and $\tau=T/T_{dec}$ with $T_{dec} \equiv 
r_{dec}/(2 \Gamma_{sh}^2 c$) have been used. $T_{dec}$ is the time-scale for 
the onset of the deceleration given in equation (\ref{gamma}). At given $T$, 
the fluid moving directly towards the observer ($\theta=0$) is located at
\begin{equation}
x_{max} = \left[ \left( 2n+1 \right) \tau \right] ^{1/\left( 2n+1 \right)} 
r_{dec} \; ,
\end{equation}
this being where the radius is 
largest and Lorentz factor is smallest on the surface, $\gamma_{sh,o} =  
\left[ \left( 2n+1 \right) \tau \right] ^{-n /\left( 2n+1 \right) }\Gamma_{sh}$.
The relativistic shock condition $\gamma_{sh} \geq 2$ used in deriving equation 
(\ref{time}) limits the applicability of equation ({\ref{equalT}) to $\tau \leq 
(2n+1)^{-1} (\Gamma_{sh}/2)^{2+1/n}$. As an example, we show in Figure 1 the 
equal-$T$ surfaces at different values of $\tau$, for $\Gamma_0=500$ 
(corresponding to $\Gamma_{sh}=330$), for a homogeneous external medium 
($\alpha=0$). The times indicated in the legend correspond to 3.6 hours, 1.5 
days, 5.0 days (and 15 days, right panel), if $T_{dec}$ is taken as 13 seconds, 
which would be the value in a burst arising from $\Gamma_0=500$, an initial 
energy release of $E_0=10^{52}\, {\rm ergs/sr}$, and a redshift $z=1$. (A jet 
of angle $\theta_{jet}$ reduces the energy required by a factor $\theta_{jet}^{-2}$
without changing the results, as long as $\gamma_{sh} \simg \theta_{jet}^{-1}$).

It is customary in analytic derivations to consider that at a
given time $T$ the emitting surface is located at 
$r = 2\gamma^2(T) cT$, and that the disk seen by the observer has a radius 
$R = r/(2\gamma(T)) = \gamma(T) cT$, as it would be in the absence of the 
deceleration (i$.$e$.$ an ellipsoid), and to calculate the properties of the 
received radiation using the physical parameters (magnetic field, electron and 
flow Lorentz factor etc$.$) of the fluid located at ($x=r,y=0$), the center of the 
projected surface. When deceleration is present, 
the radial coordinate $x_{max}$ of the center of the equal-$T$ surface can be 
related to $T$ 
by integrating $dT = dr/(2\gamma_{sh}^2 c)$ and using equation (\ref{gamma}): 
\begin{equation}
x_{max} = 4(2n+1) \gamma_o^2 c T \;,
\label{xmax}
\end{equation}
where we used the flow Lorentz factor $\gamma_o \equiv \gamma(x_{max})$ instead 
of that of the shock $\gamma_{sh,o}=\sqrt{2}\gamma_o$. Therefore $x_{max}$ is 
larger by a factor $2(2n+1)$ then the value that is typically used,
$2\gamma_o^2 cT$.  For $n=1.5$ (adiabatic remnant and homogeneous external 
gas) one obtains in equation (\ref{xmax}) a factor 16
(Sari 1997), but this factor could be as large as 28 for a strong coupling 
radiative remnant ($n=3$). This differs substantially from the usual factor 2
in the longitudinal size.  To estimate the corresponding departures
in the largest transverse size $y_{max}$ observable assuming the geometry of an
ellipsoid for the equal-$T$ surface in a decelerating fireball, we consider the
projection of this surface on the plane perpendicular to the central \los
\begin{equation}
y_{max}^{ell} = x_{max}/(2\gamma_{o}) = 2(2n+1) \gamma_{o} cT \,.
\label{ymax}
\end{equation}
The numerical factor in equation (\ref{ymax}) is 8 for $n=1.5$. 
Waxman (1997b) argued that such a large transverse size is incompatible with 
observations and that the correct transverse source size is smaller by a 
factor 4 than that inferred using a factor 16 in equation (\ref{xmax}). 
The large factor in equation (\ref{ymax}) is due to the inappropriate use of 
the geometry of an ellipsoid in the case of a decelerated expansion. 
Since the transverse source size is important for self-absorption 
considerations, for the evolution of the timescale and amplitude of afterglow 
radio scintillations and for gravitational microlensing, it is worth 
calculating accurately the coefficient in equation (\ref{ymax}) that gives 
the source size.

\section{Characteristic Radiation and Dimensions}
\label{sec:rad}

In order to determine the properties of the radiation received by the observer, 
one has to integrate over the surface of equal arrival time the emission from 
different parts of the shocked fluid, taking into account relativistic effects 
and the fact that each infinitesimal ring [$\theta,\theta+d\theta$] is 
characterized by different physical parameters (magnetic field, electron 
density, electron Lorentz factor, flow Lorentz factor). We assume that the
electrons cool only through synchrotron radiation (our numerical simulations
show that this is a good approximation),
and that they are either in the radiative or in the adiabatic
regime. In the former case we 
take into account that the remnant can be either radiative or adiabatic,
depending on the strength of the 
coupling between the radiating electrons, protons and the magnetic field (for 
more details on the radiative regime and the dynamics, see \Mesz, Rees \& 
Wijers, 1997), while in the latter case the remnant can be 
only adiabatic (energy conservation). 
Using the scaling relationships for the magnetic field 
$B' \propto \gamma r^{-\alpha/2}$, the co-moving electron density $n'_e \propto 
\gamma r^{-\alpha}$, synchrotron cooling time-scale $t'_{sy} \propto r 
\gamma^{-3} $, expansion time-scale $t'_{exp} \propto \gamma^{-1} r$, 
synchrotron power $P'_{sy} \propto \gamma^4 r^{-\alpha}$ and peak of 
synchrotron spectrum $\nu'_p \propto \gamma^3 r^{-\alpha/2}$, where $\gamma$ is 
the flow Lorentz factor, the co-moving spectral intensity at the synchrotron 
peak $I'_{\nu'_p} \propto n'_e (P'_{sy}/\nu'_p) \min{\{t'_{sy},t'_{exp}\}}$ is
\begin{equation}
I'_{\nu'_p} \propto \left\{ \begin{array}{ll}
\gamma^{-1} r^{-\alpha/2} &  {\rm for \;\; radiative \;\; electrons} \\
\gamma r^{1-3\alpha/2} &  {\rm for \;\; adiabatic \;\; electrons} \end{array} 
\right. \; .
\end{equation}
The observed spectral intensity at the detector frame peak of the synchrotron 
spectrum $\nu_p = \nu'_p /\left[\gamma\left(1-\beta \cos{\theta}\right)\right]$ 
is $I_{\nu_p} = (\nu_p/\nu'_p)^3 I'_{\nu'_p}$, therefore
\begin{equation}
I_{\nu_p} \propto \left\{ \begin{array}{ll}
a^{-(2n+\alpha/2)} (1+\gamma^2\theta^2)^{-3}  & t'_{sy} < t'_{exp} \\
a^{-(4n+1.5\alpha-1)} (1+\gamma^2\theta^2)^{-3}  &   t'_{sy} > t'_{exp} \end{array}
\right. \;\;.
\label{Ip}
\end{equation}
The bolometric co-moving intensity is $I' \sim I'_p \nu'_p$, and the observed
bolometric intensity is $I \sim  (\nu_p/\nu'_p)^4 I'$. The synchrotron spectrum 
is approximated as a broken power-law: $I_{\nu} = (\nu/\nu_p)^{\epsilon} I_{\nu_p}$ 
below the peak  ($\nu < \nu_p$) and $I_{\nu} = (\nu/\nu_p)^{-\varepsilon} 
I_{\nu_p}$ above the peak ($\nu > \nu_p$). We considered that
$\epsilon=-1/2$ and $\varepsilon = p/2$ for radiative electrons
and $\epsilon=1/3$ and  $\varepsilon = (p-1)/2$  for adiabatic electrons, $p$ being
the index of the power-law distribution of electrons (we used $p=2.5$).  The 
detector flux density is obtained by integrating $I_{\nu}$ over the equal 
arrival time surface, taking into account the solid angle subtended by each 
infinitesimal ring  [$\theta,\theta+d\theta$] on this surface. 

We study first the bolometric brightness distribution on the equal-$T$ surface. 
Figure 1 shows where most of the radiation comes from: the  upper half 
highlighted zone radiates 50\% of the total energy; 25\% of it is emitted by 
the cap extending from $\theta=0$ up to the indicated region and the other 25\% 
is radiated by the area extending toward the origin. Similarly, the lower half
highlighted part radiates 80\% of the energy received at detector. Projecting 
any of these zones on the plane perpendicular to the central \los results in 
rings that are thin compared to the disk of the entire projected surface 
seen by the observer. 

Apart from the errors mentioned at the end of \S \ref{sec:arrival}, a new
source of inaccuracy in analytic calculations of the source size and the
radiation received comes from the 
fact that the observer does not receive most of the flux from the central \los 
of the deformed ellipsoid. There is a significant difference between the 
average radial coordinates of the regions highlighted in Figure 1 and that of 
the fluid on the \los toward the center, and one must keep in mind that 
all relevant radiation parameters are power-laws in $\gamma$, which is a 
power-law in $r$. We can assess the error in equation (\ref{ymax}), obtained 
using the ``ellipsoid approximation", by calculating an intensity-weighted
average transverse coordinate $\overline{y}$ on the equal-$T$ surface and 
comparing it to $\gamma_{o} cT$. We also calculate an intensity-weighted 
average longitudinal
$\overline{x}$ coordinate on the same surface and compare it with $x_{max}$ 
from equation (\ref{xmax}), so that the factor $(x_{max}/\overline{x})^n$ will 
estimate the difference between an average $\overline{\gamma}$ that should 
in general be used instead of $\gamma_o$. These averages are given in Table 1, 
which also lists the width $w$ of the outer ring of the source projection on 
the plane perpendicular to the central \los, which contains 50\% of the 
entire flux received at detector.
Also in Table 1 are the ratios between the intensity-averaged 
synchrotron peak frequency $\overline{\nu_p}$ over the $T$-surface, and 
$\nu_o \equiv \nu_p(\theta=0)$. We find that the coefficient $2(2n+1)$ in 
equation (\ref{ymax}) over-estimates the true value of the transverse size
$\overline{y}$ by a factor $2.1-2.7$ .

We consider now observations in a fixed frequency band. Figure 2 (upper graphs) 
shows the transverse distribution of the observed synchrotron peak frequency 
$\nu_p$ for various constant-$T$ surfaces. The unknown coefficient 
in the expression for $\nu_p$ was chosen so that at $T \sim 1\; {\rm day}$
(with $T_{dec}= 13 \; {\rm s}$), most of the energy the observer receives
is in the optical range ($\sim 2$ eV).
The lower graphs show the peak (or bolometric) luminosity integrated up to the 
transverse coordinate $y$ as a function of $y$. The ring is indicated by the
steep rise in the integrated luminosity, during which $\nu_p$ varies 
by approximately one order of magnitude around the peak frequency 
$\nu_p(y_{max})$ of the radiation from the region that is seen tangentially 
by the observer. If observations are made at energies $\siml 10^{-1}
\nu_p(y_{max})$ (e$.$g$.$ in radio, for the times chosen in Figure 2) then 
the observer practically sees only the low-energy part of slope -1/2 (or 1/3) 
of the synchrotron spectrum of the radiation emitted from most points on the 
equal-$T$ surface, and the entire disk appears almost equally bright. 
However, if observations are made at energies $\simg 10\, \nu_p(y_{max})$ 
(optical or X-ray for Figure 2), then the observer sees mainly the high-energy 
tail of slope $-(p-1)/2$ (or $-p/2$) of the synchrotron spectrum from the 
power-law distribution of electrons, and the visible region reduces to a ring. 
For a given observed frequency band, as the shocked fluid is decelerated, 
$\nu_p(y_{max})$ crosses the observed band, and the region radiating in 
that band shrinks from the full disk to a narrow ring with outer boundary at
$y_{max}$, the edge of the radiating surface. During this transition 
$\overline{x}/x_{max}$ decreases while $\overline{y}/y_{max}$ increases. At 
energies far above or below $\nu_p(y_{max})$, these quantities and the width of 
the ``visible" zone are approximately constant in time. Table 2 gives the 
asymptotic range of the same coefficients as Table 1, for observations made at 
a given frequency.  The first number in each column gives the value of the 
coefficient when the source is seen as disk ($\nu \ll \overline{\nu_p} \sim 
\nu_p(y_{max})$, larger width $w$), and the last number gives the asymptotic 
value of the coefficient when the source has reduced to a ring ($\nu \gg 
\nu_p(y_{max}$), smaller $w$). The coefficients have the same range for all
frequencies. The particular frequency of the observing band only determines 
the time when the gradual transition between the two coefficients is made, 
earlier in X-rays (few hours) than in optical ($\sim$ 1 day) or radio ($\sim$ 
10 days). It can be seen that the radiative remnant gives narrower rings, and 
that the ring is wider for expansion into a decreasing density medium (e.g. 
$\alpha=2$) than into a homogeneous medium ($\alpha=0$).

\section{Discussion}
\label{sec:disc}

The main conclusions to be drawn from the calculations presented here are:\\
1. For the afterglow of a fireball, the equal arrival time surfaces are
   distorted ellipsoids whose shape and evolution depend on the dynamical regime
   of the remnant, the electron radiative efficiency and the density 
   distribution of the external medium. \\
2. Equation (\ref{xmax}) should be used to relate the observer time $T$ with
   the radial coordinate of the center of this surface, and the size of the
   source should be calculated with an equation similar to (\ref{ymax}) but 
   with the coefficient $2(2n+1)$ replaced by the coefficients in Table 1 for 
   bolometric observations. For band observations these coefficients change in
   time; Table 2 gives only their upper and lower limits. This is of
   relevance for the scintillation of the afterglow in radio (Goodman, 1997;
   Frail et al, 1997). \\
3. The spectrum and intensity should be calculated using the gas parameters
   characteristic of the ring (rather than those of the \los to the center).
   The mean ratio between average peak frequency in the ring and in the central    \los is given in Table 1. For a given dynamic and radiative regime this 
   ratio is constant in time, thus the power-law time dependences of the 
   observable flux predicted by the fireball afterglow models (\Mesz \& Rees, 
    1997; \Mesz, Rees \& Wijers, 1997) remain unchanged.\\
4. When observations are restricted to a narrow energy band, the shape
   (ring or disk) of the source seen by the observer is dependent on the
   observational band. In any band, the observer should see the source 
   increasing in overall size and changing its shape from a full disk to a 
   relatively narrow ring, at least while the expansion is relativistic.  
   This is of importance for the possible gravitational microlensing of 
   afterglows (Loeb \& Perna 1997). If bolometric observations are obtained 
   by piecing together band observations spanning many orders of magnitude, 
   then most of the energy of the afterglow should be seen coming from a 
   relatively narrow ring, at any time.

The thickness of the zone that radiates most of the energy is important 
because it determines the spread $\delta T$ in the arrival time of photons
emitted at lab-frame $t$. Generally, $\delta T$ is comparable to $T$.
In the radiative case electrons cool on a time-scale much shorter than
the expansion time, only a very thin zone located behind the blast wave front
releases significant energy and the thickness of the radiating fluid can be 
neglected. Therefore in this case the emitting fluid can be safely approximated 
as a surface.  The effect of the shell thickness is important only in the 
adiabatic case, and was taken in consideration by Waxman (1997b). We have 
employed here the equal arrival time surfaces using the kinematics of the blast 
wave. Radiation emitted by the fluid behind this shock is received at time $T > 
T_{sh}$, where $T_{sh}$ is the arrival time from the blast wave. Taking into 
account that the source size is increasing in time, it results that a finite 
thickness leads to wider rings and lower values of the averages $\overline{x}$ 
and $\overline{y}$ given in Tables 1 and 2. Our estimations of the ring's
width for an adiabatic remnant are larger by a factor up to 2.3 than calculated
by Waxman (1997b).  

Additional complications arise if different regions on the equal-$T$ surface 
are in different dynamic and/or radiative efficiency regimes. As the fireball 
decelerates, the cooling time-scale of electrons increases and they eventually 
become adiabatic. If there is a strong coupling between electrons and protons 
+ magnetic field, the remnant and electrons evolve together from the radiative 
to the adiabatic regime. In the likely case that the coupling is weak (e.g.
\Mesz, Rees, \& Wijers, 1997), the
remnant evolution is adiabatic throughout, and only the electrons evolve from
radiative to adiabatic. At times $\simg 1$ day, the most likely case is given 
by the third (or sixth) line of Tables 1 and 2. The real situation may be even 
more complex if the power-law electrons are in different radiative regimes,
e.g. low energy electrons may be adiabatic ($t'_{sy} > t'_{exp}$) while 
higher energy electrons may be radiative ($t'_{sy} < t'_{exp}$).

\acknowledgements{This research has been supported by NASA NAG5-2857 and NAG5-2362}

\begin{table}
\begin{center}
TABLE 1. \\ {\sc Intensity-averaged parameters on the equal arriving time surface \\ 
            and the width $w$ of the ring seen  by the observer for bolometric observations} \\ [4ex]
\begin{tabular}{cccccccc} \hline
\rule[-4mm]{0mm}{10mm} n [$\alpha,\delta$]  & 2(2n+1) & $\overline{x}/x_{max}$ & $\overline{x}/(2\gamma_{o}^2 cT)$ 
   & $\overline{y}/y_{max}$  & $\overline{y}/(\gamma_{o} cT)$  & $\overline{\nu_p}/\nu_{o}$ &  $w$  \\ \hline
\rule[-2mm]{0mm}{6mm} $3.0\;[0,0]^r$ & 14  &  0.78  &  11  &  0.87  &  5.2  &  39   &  0.07  \\
\rule[-2mm]{0mm}{6mm} $1.5\;[0,1]^r$ &  8  &  0.82  &  6.6 &  0.76  &  3.2  &  4.0  &  0.17  \\
\rule[-2mm]{0mm}{6mm} $1.5\;[0,1]^a$ &  8  &  0.72  &  5.8 &  0.82  &  3.4  &  8.3  &  0.11  \\ \hline
\rule[-2mm]{0mm}{6mm} $1.0\;[2,0]^r$ &  6  &  0.74  &  4.4 &  0.77  &  2.6  &  6.7  &  0.16  \\
\rule[-2mm]{0mm}{6mm} $0.5\;[2,1]^r$ &  4  &  0.78  &  3.1 &  0.71  &  1.8  &  2.0  &  0.23  \\
\rule[-2mm]{0mm}{6mm} $0.5\;[2,1]^a$ &  4  &  0.49  &  2.0 &  0.76  &  1.9  &  6.8  &  0.20  \\ \hline
\end{tabular}
\begin{displaymath} 
^r\; {\rm radiative\; electrons} \quad ^a\; {\rm adiabatic\; electrons} 
\end{displaymath}
\end{center}
\end{table}

\begin{table}
\begin{center}
TABLE 2. \\ {\sc Intervals for intensity-averaged parameters on the equal arriving time surface \\ 
            and width $w$ of the ring seen by the observer for band observations} \\ [4ex]
\begin{tabular}{cccccc} \hline
\rule[-4mm]{0mm}{10mm} n [$\alpha,\delta$]  & $\overline{x}/x_{max}$ & $\overline{x}/(2\gamma_{o}^2 cT)$
                        & $\overline{y}/y_{max}$  & $\overline{y}/(\gamma_{o} cT)$  &  $w$  \\ \hline
\rule[-2mm]{0mm}{6mm} $3.0\;[0,0]^r$ &  0.88--0.72  &  12--10  &  0.77--0.89  &  4.6--5.4  &  0.16--0.06 \\
\rule[-2mm]{0mm}{6mm} $1.5\;[0,1]^r$ &  0.87--0.78  & 7.0--6.3 &  0.70--0.79  &  2.9--3.3  &  0.25--0.13 \\
\rule[-2mm]{0mm}{6mm} $1.5\;[0,1]^a$ &  0.89--0.77  & 7.1--6.2 &  0.67--0.79  &  2.8--3.3  &  0.30--0.14 \\ \hline
\rule[-2mm]{0mm}{6mm} $1.0\;[2,0]^r$ &  0.83--0.68  & 5.0--4.1 &  0.72--0.80  &  2.4--2.7  &  0.23--0.13 \\
\rule[-2mm]{0mm}{6mm} $0.5\;[2,1]^r$ &  0.83--0.75  & 3.3--3.0 &  0.68--0.73  &  1.7--1.8  &  0.28--0.21 \\
\rule[-2mm]{0mm}{6mm} $0.5\;[2,1]^a$ &  0.78--0.56  & 3.1--2.3 &  0.70--0.76  &  1.7--1.9  &  0.25--0.20 \\ \hline
\end{tabular}
\begin{displaymath}
^r\; {\rm radiative\; electrons} \quad ^a\; {\rm adiabatic\; electrons}
\end{displaymath}
\end{center}
\end{table}


\clearpage

\begin{figure}
\centerline{\psfig{figure=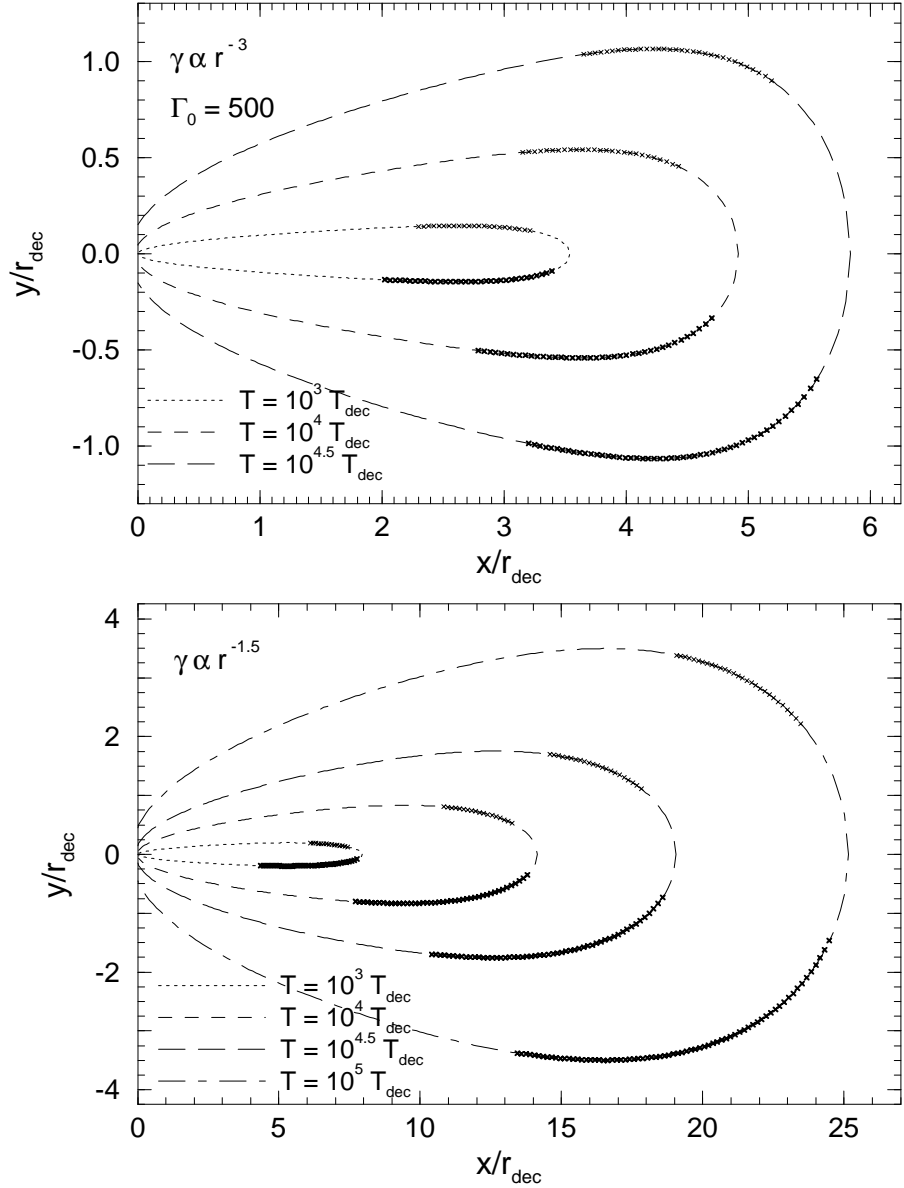}}
\vspace*{3in}
\figcaption{Surfaces of equal arrival times, for a homogeneous external medium
and a radiative (top) or adiabatic remnant dynamics (bottom).
Each curve is a transverse section through the 3-dimensional equal-$T$
surface, highlighting the regions that radiate 50\% (upper half of each curve)
and 80\% (lower half) of the bolometric flux. Projected on the plane
perpendicular to the \los toward the center of explosion, these regions appear
narrower than the projection of the entire radiating surface. The
Cartesian coordinates are normalized to $r_{dec}$, which for the putative
burst parameters in the text is 
$\sim 4\times 10^{16}$ cm, corresponding at
a redshift $z=1$  ($H_0= 75\, {\rm km\,s^{-1} Mpc^{-1}}$, $\Omega=1$) to an
angular scale $2.5\,{\rm \mu as}$.}
\end{figure}

\clearpage

\begin{figure}
\centerline{\psfig{figure=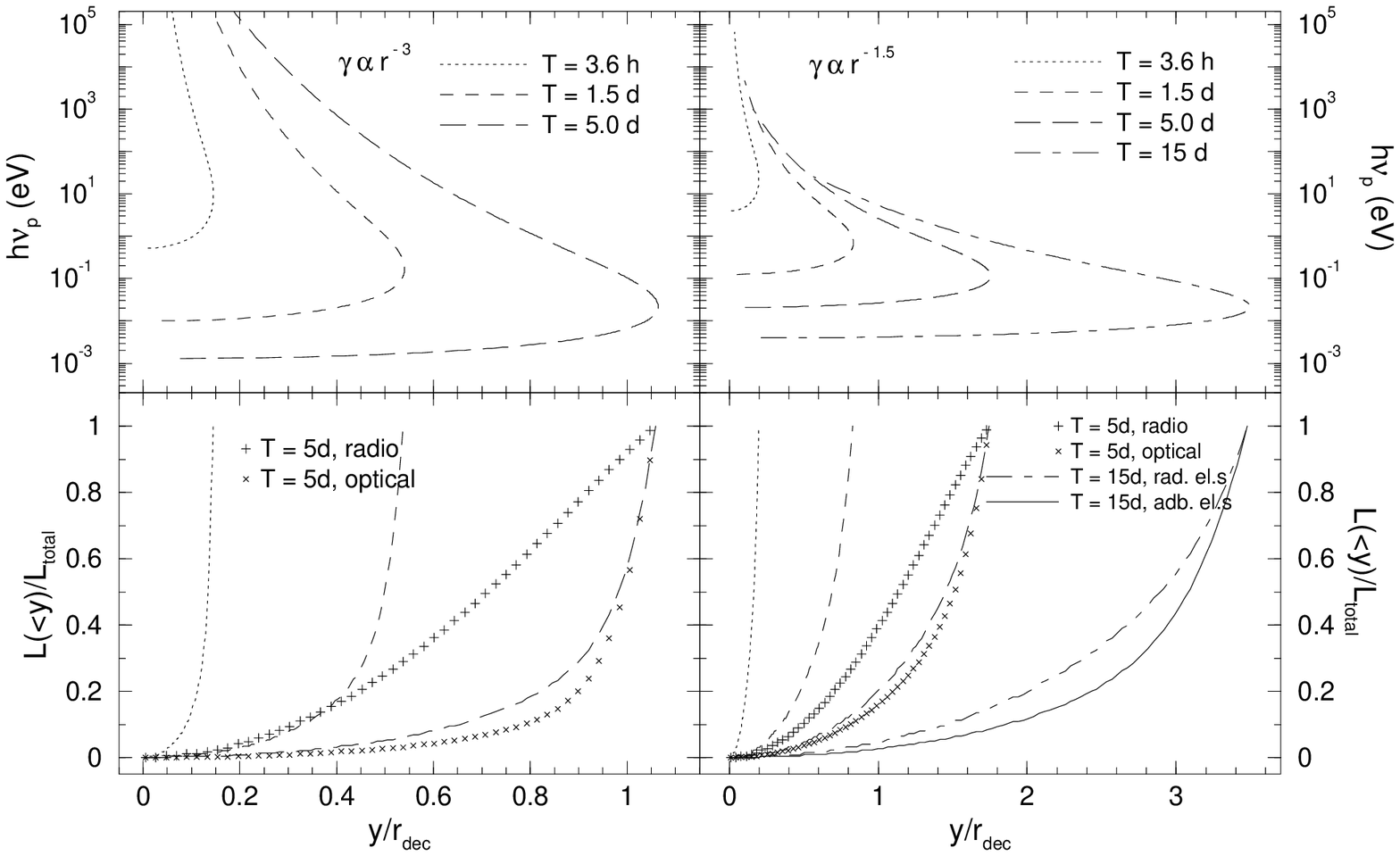}}
\vspace*{1in}
\figcaption{Distribution of peak of synchrotron spectrum on the equal
arriving time surface (upper graphs) and bolometric luminosity on the
same surface (lower graphs). Left panels are for a radiative remnant
($n=3$), right panels for an adiabatic one ($n=1.5$). The ring is shown
by the steep rise in integrated bolometric luminosity in the lower graphs.
Also shown in the lower graphs are the widths as observed in two fixed 
frequency bands (2 eV and 10 GHz). The lower right graph shows that if 
electrons are adiabatic (solid curve), then the bright zone shrinks to 
an even narrower ring.}
\end{figure}


\begin{thebibliography}{}

\bibitem[]{} Frail D., et.al., 1997, Nature, in press
\bibitem[]{} Goodman, J. 1997, New Astr., submitted (astro-ph/9706084)
\bibitem[]{} Loeb, A. \& Perna, R. 1997, ApJL, submitted (astro-ph/9708159)
\bibitem[]{} \Mesz, P. \& Rees, M. J. 1997, ApJ, 476, 232
\bibitem[]{} \Mesz, P., Rees, M. J. \& Wijers, R., 1997, ApJ, submitted (astro-ph/9709273)
\bibitem[]{} Panaitescu, A. \& \Mesz, P., 1997, ApJ, 492, in press (astro-ph/9703187)
\bibitem[]{} Rees, M. J. 1966, Nature, 211, 468
\bibitem[]{} Sari, R. 1997, astro-ph/9706078
\bibitem[]{} Tavani, M., 1997, ApJL, 483, L87
\bibitem[]{} Vietri, M., 1997, ApJ, submitted (astro-ph/9706060)
\bibitem[]{} Waxman, E., 1997a, ApJL, 485, L5
\bibitem[]{} Waxman, E. 1997b, ApJL, submitted (astro-ph/9709190)
\bibitem[]{} Wijers, R.A.M.J., Rees, M.J. \& \Mesz, P., 1997, MNRAS, 288, L51

\end{thebibliography}
\end{document}